\documentstyle[12pt]{article}
\input amssym.def
\input amssym.tex
\begin{document}
\hfill {ITP-96-67}
 
\hfill {DAMTP R96/37}\\
 
\begin{center}
{\LARGE\bf Nucleating Black Holes via \\
\vspace*{0.3cm}

Non-Orientable Instantons}\\
\vspace*{0.5cm}

{\large Andrew Chamblin$^{1}$ and G.W. Gibbons$^{2}$}\\
\vspace*{0.2cm}

{\small\it $^1$Institute for Theoretical Physics,}\\
{\small\it University of California,}\\
{\small\it Santa Barbara, California 93106--4030, U.S.A.}\\
\vspace*{0.5cm}

{\small\it $^2$Department of Applied Mathematics and
Theoretical
Physics,}\\
{\small\it University of Cambridge, Cambridge CB3 9EW,
England}\\
\end{center}
\vspace*{0.4cm}

{\noindent \small {\bf Abstract.} We extend the analysis of black hole pair creation
to include non-orientable instantons. We classify these instantons in terms
of their fundamental symmetries and orientations. Many of these
instantons admit the pin structure which corresponds to the fermions actually observed
in nature, and so the natural objection that these manifolds do not admit spin structure may
not be relevant. Furthermore, we analyse the thermodynamical properties of non-orientable
black holes and find that in the non-extreme case, there are interesting modifications
of the usual formulae for temperature and entropy.}\\
\vspace*{0.2cm}

{\noindent \bf Introduction}\\

{Recently, there has been considerable interest in the application of
semi-classical Euclidean quantum gravity techniques to the study of black hole pair
creation. Here, the analogy is with ordinary electron-positron pair production, where
one computes the probability for the process by looking at the action of the `Wick
rotated' solution, which is an electron moving on a circle in a uniform field on
Euclidean space. Likewise, in Euclidean quantum gravity, one models generic tunnelling
phenomena by first finding an instanton (a manifold $M$, Riemannian metric $g$, and
matter fields $\{{\phi}\}$ which solve the relevant field equations) and then matching
the instanton to a Lorentzian solution across a three-surface ${\Sigma}$ of vanishing
extrinsic curvature. The amplitude for such a decay process is then given 
by $e^{-S}$, where $S$ is the Euclidean action, i.e.,}
\[
S = -\,\frac{1}{16\pi G}\,\int_{M} (R - 2{\Lambda}) \sqrt{g} \,d^{4}x - \int_{M} {\cal
L}_{m} \sqrt{g} \,d^{4}x - \frac{1}{8\pi G} \int_{\partial M} (K - K^{0}) \sqrt{h}
\,d^{3}x
\]
{where $G$ is Newton's constant, $g$ is the determinant of the four-metric, $h$ is the
determinant of the three-metric on ${\partial}M$ (which we assume is positive
definite), ${\cal L}_{m}$ is the Lagrangian of any matter fields, $R$ is the scalar
curvature of the four-manifold $M$, ${\Lambda}$ is the cosmological constant, $K$ is
the trace of the second fundamental form of the boundary (relative to the metric
$g_{ab}$ on $M$) and $K^{0}$ is the trace of the second fundamental form of the
boundary imbedded in flat space.

Instantons (with or without matter fields)
which correspond to a black hole moving on a loop in Euclidean space have the
topology $S^{2} \,\times\, S^{2}$. In the simplest case (without matter fields), one
takes the product metric, $g_{R}$, given by the direct sum of the two round metrics on
each of the $S^{2}$ factors. The nucleation surface is then the ${\Sigma} \,\simeq\, 
S^{1} \,\times\, S^{2}$, with vanishing extrinsic curvature. Thus, this metric (known
as the Nariai instanton) `nucleates' a wormhole $S^{1} \,\times\, S^{2}$. Given the
presence of horizons in the Lorentzian section, one can think of this instanton as
modelling black hole pair production in a De Sitter background, as was first noted
in [3], [4].

It has been argued [1] that the only instantons which are of any real interest, to
black hole pair creation, are those which are 
simply connected and admit a spin structure. However, this restriction may be
physically too severe.

As an example, take the case of the Nariai instanton, and identify the
`spacelike' $S^{2}$ under the antipodal map, to obtain an instanton with
topology $S^{2} \,\times\, {\Bbb {RP}}^{2}$ and nucleation surface ${\Sigma} \,\simeq\,
S^{1} \,\times\, {\Bbb {RP}}^{2}$. The Lorentzian section of this solution
contains `black holes' (which are now {\it non-orientable}), and so it corresponds to
the birth of a non-orientable wormhole. Because $w_{2}({\Bbb {RP}}^{2})
\,\neq\, 0$, the instanton does not admit a spin structure. As we shall show, however,
it {\it does} admit the pin structure observed in nature (more precisely, 
$S^{2} \,\times\, {\Bbb {RP}}^{2}$ admits the pin structure which particle physicists
customarily use to construct the discrete maps $P$ and $T$ on the Hilbert space of
solutions to the Dirac equation [5]). Thus, we can see no reason to exclude this
creation process from consideration. On the contrary, the thesis of this work is that 
if one
accepts the semi-classical approach as a valid approximation to ordinary black hole
pair creation, then one must also accept the pair creation of non-orientable black
holes.

Historically ([6], [7]), the subject of non-orientable black holes has been rather
neglected since of course a non-orientable hole would never form in a
realistic astrophysical scenario involving gravitational
collapse . Now that we have a physical mechanism for creating such objects,
there will hopefully be a renewal of interest. On a more philosophical note, we feel
that these issues are important because they focus attention on more `exotic'
manifolds which ordinarily are overlooked by those studying the quantum
foam. After all, a {\it truly} robust implementation of Feynman's ideas to gravity
would require that we sum over {\it all manifolds} first, 
and determine afterwards which contributions may vanish, for example
because of a vanishing fermionic determinant factor on an infinite bosonic
determinant factor.}\\
\vspace*{0.6cm}

{\noindent \bf I. Schwarzschild and the Elliptic Interpretation}\\

{In this section, we recall the
properties of `classical' non-orientable black holes, as described previously in [7].
To this end, let $({\cal M}, g)$ denote the Schwarzschild spacetime. We are interested
in identifying ${\cal M}$ under the action of certain discrete involutive isometries.
In particular, we are interested in the actions of time and space inversion. Since we
wish to consider the actions of these inversions on the maximally extended spacetime,
it is most natural to use Kruskal coordinates [12], which cover the entire manifold. For
reasons which will become apparent, we feel it is useful to first review the relation
of these coordinates to the usual Schwarzschild coordinates (which cover only part of
the maximal extension).

As usual, let  $(t, r, {\theta}, {\phi})$ denote the Schwarzschild coordinates
so that the metric reads}
\[
ds^{2} = -\left(1 - \frac{2m}{r}\right)dt^{2} \,+\, \frac{dr^{2}}{\left(1 - 
\frac{2m}{r}\right)} \,+\, r^{2}\left(d{\theta}^{2} \,+\,
{\sin}^{2}{\theta}d{\phi}^{2}\right) .
\]
{Next introduce null coordinates $u$ and $v$ such that}
\begin{eqnarray*}
du = dt - \frac{dr}{\left(1 - \frac{2m}{r}\right)} ,\\
 \\
dv = dt + \frac{dr}{\left(1 - \frac{2m}{r}\right)} ,
\end{eqnarray*}
{or, integrating}
\begin{eqnarray*}
u = t -r -2m \log (r - 2m) ,\\
v = t + r + 2m \log (r - 2m) .
\end{eqnarray*}
{Now form the coordinates $U$ and $V$ by exponentiating}
\begin{eqnarray*}
U &=& - e^{\frac{- u}{4m}} ,\\
V &=& e^{\frac{- v}{4m}} ,
\end{eqnarray*}
{Then one finds that the coordinates $T$ and $Z$ defined by}
\begin{eqnarray*}
T = \sinh \left(\frac{t}{4m}\right) e^{\frac{r}{4m}} \sqrt{r - 2m} \\
 \\
Z = \cosh \left(\frac{t}{4m}\right) e^{\frac{r}{4m}} \sqrt{r - 2m}
\end{eqnarray*}
{satisfy the simple algebraic relations}
\begin{eqnarray*}
U = T \,+\, Z ,\\
V = T \,-\, Z ,
\end{eqnarray*}
{that is, $U$ and $V$ are advanced and retarded null coordinates relative to $T$ and
$Z$. One checks that in these coordinates the metric assumes the form}
\[
ds^{2} = e^{\frac{-r}{2m}}\left(16\frac{m^{2}}{r}\right)\left(-dT^{2} \,+\,
dZ^{2}\right) \,+\,
r^{2}\left(d{\theta}^{2} \,+\, {\sin}^{2}{\theta}\,d{\phi}^{2}\right) .
\]
{Using the coordinates $(T, Z, {\theta}, {\phi})$, we define total time
inversion by the map}
\[
{R_{T}: \,(T, Z, {\theta}, {\phi}) \,\longrightarrow\, (-T, Z, {\theta}, {\phi})} ,
\]
{and likewise space inversion is given by}
\[
{R_{Z}: \,(T, Z, {\theta}, {\phi}) \,\longrightarrow\, (T, -Z, {\theta}, {\phi})} .
\]
{Of course, neither of these involutions acts freely (they both have fixed points). To
obtain a free action, we need to take a product with some other map which {\it is}
freely acting. Such a map, which we denote as `$P$', is given as follows:}
\[
{P: \,(T, Z, {\theta}, {\phi}) \,\longrightarrow\, (T, Z, {\pi} - {\theta}, 
{\phi} + {\pi})}
\]

{Thus, we can construct the following four freely acting involutions on ${\cal M}$:
$P, PR_{T}, PR_{Z}$ and $PR_{Z}R_{T}$. We claim that all of these involutions extend
to the corresponding Euclidean instanton (the `cigar'). Before addressing the
Riemannian issue, however, we need to first define and interpret the basic properties
of the spacetime obtained when we identify ${\cal M}$ under the action of one of these
involutions. To this end, let $J$ be any one of the above involutions. We want
to consider the {\it quotient} manifold}
\[
{\cal M}_{J} = {\cal M}/J
\]
{Depending on which choice we make for $J$, ${{\cal M}_{J}}$ may or
may not be asymptotically flat and it may or may not be orientable. However, a little
thought establishes the following table:}\\
\begin{center}
\begin{tabular}{|l|c|c|c|}
\hline
 & & & \\
 &${{\cal M}_{J}}$ &${{\cal M}_{J}}$ 
&${{\cal M}_{J}}$ \\
\hspace*{1.5cm} $J$ &asymptotically &time &space \\
 &flat? &orientable? &orientable? \\
 & & & \\
\hline
 & & & \\
Case I: $J = R_{T}R_{Z}P$ &yes &no &yes \\
 & & & \\
\hline
 & & & \\
Case II: $J = R_{T}P$ &no &no &no \\
 & & & \\
\hline
 & & & \\
Case III: $J = R_{Z}P$ &yes &yes &yes \\
 & & & \\
\hline
 & & & \\
Case IV: $J = P$ &no &yes &no \\
 & & & \\
\hline
\end{tabular}
\end{center}
\begin{center}
{\large Table 1}
\end{center}
\vspace*{0.2cm}

{\noindent Thus, we see that the only `nice' quotient manifold (i.e., the only 
one which is both
asymptotically flat and orientable) is ${{\cal M}_{ZP}} = {\cal M}/{J}$,
with $J = R_{Z}P$. (Note: We will employ this notation from here on, i.e., 
${{\cal M}_{TZP}}$ denotes ${\cal M}/{J}$ with ${J} = R_{T}R_{Z}P$, 
${{\cal M}_{TP}}$ denotes ${\cal M}/{J}$ with ${J} = R_{T}P$, and so
on). Of course, our point of view is that we should, to begin with at least, consider
all of these spacetimes on an equal footing, and not let lack of an orientation
dissuade us from studying them (although as we will point out later, a lack of time
orientation would seem to be a problem when one introduces quantum mechanics). To this
end, consider the spacetime ${{\cal M}_{P}}$.

Although this quotient manifold is not asymptotically flat (the antipodal
identification forces the spacelike slices to have the wrong topology, i.e., ${\Bbb
R}{\Bbb P}^{2} \,\times\, [0, 1)$, at large spatial distances), we consider it anyway
since in the context of cosmological pair creation, the background in which the 
black holes are produced is not asymptotically flat. 

Another natural question about ${{\cal M}_{P}}$ is whether or not
one can `tell' that it is non-orientable. What effect would a lack of
space-orientation have? For example, could we use an object such as the black hole
described by ${{\cal M}_{P}}$ to turn right-handed people into left-handed
people (and vice versa)? Clearly, the answer to this question is yes,
since by simply moving around the perimeter of the hole an odd number of times
we traverse a non-trivial generator of ${\pi}_{1}({\Bbb R}{\Bbb P}^{2})$, and
such a curve is by definition a space-orientation reversing curve in 
${{\cal M}_{P}}$ (although the space-time curve corresponding to such  
causal movement is not closed, it is homotopic to a closed curve in the 
spatial factor).

Interesting questions can also arise when one considers the inclusion of
quantum effects on ${{\cal M}_{P}}$. For example, the area of the event horizon
in ${{\cal M}_{P}}$ (which has topology ${\Bbb R}{\Bbb P}^{2}$) is
$\frac{1}{2}$ times the area of the horizon in ${\cal M}$ (with topology $S^{2}$), the
reason being that one way to calculate the area of a non-orientable surface is to
calculate the area of its double-cover and divide by two. Will this discrepancy in
areas affect the temperature? In order to answer this question properly, we
need to look carefully at the corresponding Riemannian instanton, the `identified
cigar'. To this end, and also to see how the freely acting involutive isometries on
the Lorentzian section are related to freely acting involutive isometries
on the Riemannian section, let us write complexified Schwarzschild as an algebraic
variety in ${\Bbb C}^{7}$ as usual. More explicitly, let $\{Z^{i}\,|\,i = 1,\, 
\dots \, 7\}$ be coordinates on ${\Bbb C}^{7}$, so that in terms of Schwarzschild
coordinates (which cover only a subset of the variety), we have [8]}
\begin{eqnarray}
Z^{1} = r\sin \theta \cos \phi , \nonumber \\
\nonumber \\ 
Z^{2} = r\sin \theta \sin \phi , \nonumber \\
\nonumber \\ 
Z^{3} = r\cos \theta ~~~~~~ , \nonumber \\
\nonumber \\
Z^{4} = -2M \sqrt{\frac{2M}{r}} \,+\, 4M \sqrt{\frac{r}{2M}} ,\\
\nonumber \\
Z^{5} = 2M \sqrt{3} \sqrt{\frac{2M}{r}}~~~~~~ , \nonumber \\
\nonumber \\
Z^{6} = 4M \sqrt{1 - \frac{2M}{r}} \cosh \left(\frac{t}{4M}\right) , \nonumber \\
\nonumber \\
Z^{7} = 4M \sqrt{1 - \frac{2M}{r}} \sinh \left(\frac{t}{4M}\right) . \nonumber 
\end{eqnarray}

{With the coordinates as in (1), it turns out that complexified Schwarzschild (${\cal
M_{\Bbb C}}$) is given as the algebraic variety determined by the three polynomials}
\begin{eqnarray}
(Z^{6})^{2} - (Z^{7})^{2} \,+\, \frac{4}{3} (Z^{5})^{2} = 16M^{2} , \nonumber \\
\nonumber \\
\left((Z^{1})^{2} \,+\, (Z^{2})^{2} \,+\, (Z^{3})^{2}\right) (Z^{5})^{4} = 576M^{6} , \\
\nonumber \\
\sqrt{3} Z^{4} Z^{5} \,+\, (Z^{5})^{2} = 24M^{2} . \nonumber
\end{eqnarray}

{The Lorentzian section (${\cal M} = {\cal M}^{L}$) and the Riemannian section (${\cal
M}^{R}$) are then specified by finding certain anti-holomorphic involutions acting on
the above variety which stabilise either ${\cal M}^{L}$ or ${\cal M}^{R}$; that is, we
find maps}
\begin{eqnarray*}
{J}_{L}: \,{\cal M}_{\Bbb C} \,\longrightarrow\, {\cal M}_{\Bbb C} , \\
{J}_{R}: \,{\cal M}_{\Bbb C} \,\longrightarrow\, {\cal M}_{\Bbb C} ,
\end{eqnarray*}
{such that ${J}_{L}$ leaves ${\cal M}^{L} \,\subset\, {\cal M}_{\Bbb C}$ 
invariant:}
\[
{J}_{L} ({\cal M}^{L}) = {\cal M}^{L} ,
\]
{and such that ${J}_{R}$ leaves ${\cal M}^{R} \,\subset\, {\cal M}_{\Bbb C}$
invariant:}
\[
{J}_{R} ({\cal M}^{R}) = {\cal M}^{R} .
\]

{As described in [8] ${J}_{L}$ restricted to ${\cal M}^{L}$ is an anti-holomorphic
version of time reversal. ${J}_{R}$ is the map given by reflection through the
${\tau} = 0$ (where ${\tau} = it$) three-surface in the `cigar' instanton (i.e., 
${\tau} = 0$ is the `Einstein Rosen bridge' three-surface ${\Sigma}$, with topology
$S^{2} \,\times\, {\Bbb R}$). Since the surfaces $t = 0$ and ${\tau} = it = 0$
correspond to the surface $Z^{7} = 0$, we see that ${\cal M}^{L}$ and ${\cal M}^{R}$
intersect precisely along this Einstein Rosen bridge. Explicitly, we can realise the
two maps ${J}_{L}$ and ${J}_{R}$ as follows:}
\begin{eqnarray*}
{J}_{L}: \, (Z^{1}, Z^{2}, Z^{3}, Z^{4}, Z^{5}, Z^{6}, Z^{7}) \,\longrightarrow\, 
({\bar Z^{1}}, {\bar Z^{2}}, {\bar Z^{3}}, {\bar Z^{4}}, {\bar Z^{5}}, 
{\bar Z^{6}}, {\bar Z^{7}}) , \\
 \\
{J}_{R}: \, (Z^{1}, Z^{2}, Z^{3}, Z^{4}, Z^{5}, Z^{6}, Z^{7}) \,\longrightarrow\, 
({\bar Z^{1}}, {\bar Z^{2}}, {\bar Z^{3}}, {\bar Z^{4}}, {\bar Z^{5}}, 
{\bar Z^{6}}, - {\bar Z^{7}}) .
\end{eqnarray*}
{Comparing these explicit formulae for ${J}_{L}$ and ${J}_{R}$ with the
coordinates in (1), we see that ${J}_{R}$ is thus obtained from ${J}_{L}$ by
the transformation $t \,\longrightarrow\, {\tau} = it$.  

What we want to do now is show how the maps $R_{T}, R_{Z}$ and $P$ acting on ${\cal
M}^{L}$, and likewise their Euclidean counterparts acting on ${\cal M}^{R}$, are
actually just the {\it restrictions} to ${\cal M}^{L}$ and ${\cal M}^{R}$ of certain
{\it holomorphic} involutions acting on ${\cal M}_{\Bbb C}$. Of course, once we notice
that our complex coordinates $Z^{6}$ and $Z^{7}$ are (up to a scaling) actually our
Kruskal coordinates $Z$ and $T$, it is easy to see that the `big' involutions, ${\cal
R}_{Z}$ and ${\cal R}_{T}$ (which restrict to $R_{Z}$ and $R_{T}$ on ${\cal M}^{L}$)
are given by}
\begin{eqnarray*}
{\cal R}_{Z}: \, (Z^{1}, Z^{2}, Z^{3}, Z^{4}, Z^{5}, Z^{6}, Z^{7}) \,\longrightarrow\, 
({Z^{1}}, {Z^{2}}, {Z^{3}}, {Z^{4}}, {Z^{5}}, - {Z^{6}}, {Z^{7}}) , \\
 \\
{\cal R}_{T}: \, (Z^{1}, Z^{2}, Z^{3}, Z^{4}, Z^{5}, Z^{6}, Z^{7}) \,\longrightarrow\, 
({Z^{1}}, {Z^{2}}, {Z^{3}}, {Z^{4}}, {Z^{5}}, {Z^{6}}, - {Z^{7}}) .
\end{eqnarray*}
{Clearly, these maps are holomorphic, and since they commute with both $J_{L}$ and
$J_{R}$, they restrict to well-defined involutions on ${\cal M}^{L}$ and 
${\cal M}^{R}$. Thus, ${\cal R}_{Z}|_{{\cal M}^{L}} = R_{Z}$ and 
${\cal R}_{T}|_{{\cal M}^{L}} = R_{T}$. For the maps restricted to the Riemannian
section, we shall write}
\begin{eqnarray*}
{\cal R}_{Z}|_{{\cal M}^{R}} = {\bar R_{Z}}: \, {\cal M}^{R} \,\longrightarrow\, {\cal
M}^{R} \\
 \\
{\cal R}_{T}|_{{\cal M}^{R}} = {\bar R_{T}}: \, {\cal M}^{R} \,\longrightarrow\, {\cal
M}^{R} 
\end{eqnarray*}
{In terms of local coordinates on ${\cal M}^{R}$, these reflections take the form}
\begin{eqnarray*}
{\bar R_{Z}}: \, {\tau} \,\longrightarrow\,  -{\tau} \,+\, 4{\pi}m \\
{\bar R_{T}}: \, {\tau} \,\longrightarrow\, - {\tau}
\end{eqnarray*}
{($r, {\theta}$, and ${\phi}$ are left invariant by both these maps). Thus, we see
that ${\bar R_{T}}$ is reflection in imaginary time whereas ${\bar R_{Z}}$ corresponds
to rotating through half a period in imaginary time.

Finally, we obtain the involution ${\bar P}$ on ${\cal M}^{R}$ by
restricting to ${\cal M}^{R}$ the following map on ${\cal M}^{\Bbb C}$:}
\[
{\cal P}: \, (Z^{1}, Z^{2}, Z^{3}, Z^{4}, Z^{5}, Z^{6}, Z^{7}) \,\longrightarrow\, 
(- {Z^{1}}, - {Z^{2}}, - {Z^{3}}, {Z^{4}}, {Z^{5}}, {Z^{6}}, {Z^{7}})
\]

{Now that we have made sense of how to extend our discrete isometries $R_{Z}, R_{T}$
and $P$ from ${\cal M}$ to ${\cal M}^{R}$, we can return to the problem of examining
the thermodynamical properties of ${\cal M}_{J} = {\cal M}^{L}/J$ by looking at the
instanton ${\cal M}_{J}^{R} = {\cal M}^{R}/{\bar J}$. In particular, let us focus
again on ${\cal M}_{P}$.

First of all, since we have only identified under the action of parity inversion, we
have not affected the period of the imaginary time coordinate ${\tau}$. Na{\"\i}vely, we
might therefore expect that the temperature of the hole in ${\cal M}_{P}$ would be the
same as that in ${\cal M}$, given the thermodynamical principle [20] that
the temperature $T$ is inversely related to the period ${\beta}$:}
\[
T = {\beta}^{-1}
\]
{Indeed, this reasoning is correct and the temperature of ${\cal M}$ is in fact equal
to the temperature of ${\cal M}_{P}$; however, there are many subtleties which now
arise and one finds that in order to maintain this relation between temperature and
period, one has to alter the standard formulae which express the relations between
temperature, mass and area.

To see how this works, recall first that ${\cal M}_{P}$ is not asymptotically flat
since at large radial distances spacelike slices have the topology ${\Bbb R}{\Bbb
P}^{2} \,\times\, [0, 1)$. Let ${\cal M}_{P}$ be obtained from ${\cal M}$ under the
action of $P$, so that ${\cal M}$ is the double cover of ${\cal M}_{P}$. Then the
horizon in ${\cal M}$ is an $S^{2}$ which is the double cover of the horizon in ${\cal
M}_{P}$, which is an ${\Bbb R}{\Bbb P}^{2}$. It follows that the area, $A$, of the
horizon in ${\cal M}$ is twice the area, $A_{P}$, of the horizon in ${\cal M}_{P}$:}
\[
A = 2 A_{P}
\]

{In a similar way, we can calculate the relationship of the ADM mass, $m_{P}$, of the
hole in ${\cal M}_{P}$ to the ADM mass, $m$, of the hole in ${\cal M}$. Of course, one
might well wonder how we expect to define mass given a lack of asymptotic flatness as
it is usually understood. We posit that it still makes sense to define the mass as a
surface integral of some flux density over, a two-surface `at infinity', even if the
two-surface has the topology of ${\Bbb R}{\Bbb P}^{2}$, so that the calculation of the
mass reduces to calculating the mass of the cover and dividing by 2:}
\[
m = 2 m_{P}
\]

{Using $dm = {\kappa}dA$ we see that consistency requires that ${\kappa} = {\kappa}_{P}$
if and only if $T = T_{P}$.}

{However, now recall the formula relating the area $A$ and ADM mass $m$ in
${\cal M}$:}
\[
A = 16\pi m^{2}
\]
{Substituting the above formulae for $m$ and $A$ into this expression, we obtain}
\[
2A_{P} = 16\pi (4m_{P}^{2})
\]
{hence}
\[
A_{P} = 32\pi m_{P}^{2}
\]
{and so the usual relationship between horizon area and ADM mass is slightly changed in 
${\cal M}_{P}$.

What about temperature and entropy? Well, as we have seen above, the temperature,
$T_{P}$, of ${\cal M}_{P}$ must equal the temperature, $T$, of ${\cal M}$ since the
periods are the same:}
\[
T = T_{P}
\]
{But $T = \frac{1}{8\pi m}$, and so the usual relationship between $T$ and $m$ on
${\cal M}$ is modified on ${\cal M}_{P}$ to}
\[
T_{P} = \frac{1}{16\pi m_{P}}
\]

{\noindent \bf II.  Non-orientability and Wormholes}\\

{In this section, we want to point out that if the nucleation surface ${\Sigma}$ is
closed and  non-orientable, then $b_{1}({\Sigma})$, the first Betti
number, cannot vanish. This means that the fundamental group ${\pi}_{1}({\Sigma})$ must
contain elements of infinite order, or put more colloquially, ${\Sigma}$ must contain
Wheeler wormholes. This is clearly the case for the Nariai solution for which ${\Sigma}
\,{\cong}\, S^{1} \,{\times}\, {\Bbb R}{\Bbb P}^{2}$. The point we wish to make is that
this is always so. Note however that the element of infinite order whose existence is
ensured does not necessarily reverse orientation. That is, the wormhole we must always
have is not necessarily an orientation reversing wormhole.

The proof of this result is given in [24] and amounts to the observation that the Euler
characteristic of any three-manifold, orientable or not, vanishes, thus}
\[
{\chi}({\Sigma}) = b_{0}({\Sigma}) - b_{1}({\Sigma}) \,+\, b_{2}({\Sigma}) - 
b_{3}({\Sigma}) = 0
\]
{If ${\Sigma}$ is connected, then $b_{0}({\Sigma}) = 1$ and if ${\Sigma}$ is {\it not}
orientable, then $b_{3}({\Sigma}) = 0$. Thus}
\[
b_{1}({\Sigma}) = 1 \,+\, b_{2}({\Sigma}) \,{\geq}\, 1 .
\]
{Note that the result is false if ${\Sigma}$ has dimension greater than 3.}\\
\vspace*{0.6cm}

{\noindent \bf III.  Fermions on Non-Orientable Spacetimes}\\

{It is often stated in the literature that since it is impossible to 
define a spin structure on a non-orientable spacetime, it is impossible
to define fermions on such spacetimes.  We will now show that it is often
possible to have fermions regardless of whether or not there exists a 
spin structure.  We would also like to emphasize now that these are 
ordinary fermions, i.e., particles acted upon by the full inhomogeneous
Lorentz group.  We will not consider enlarging the group of symmetries
by coupling to some internal gauge group, as is done when one passes
to a $Spin_{c}$ structure, since as has been pointed out elsewhere [15]
such an enlargement would not correspond to the observed couplings 
between gauge bosons and fermions.  For related reading we refer the 
reader to ([13], [14]).  

Just to be concrete, let us begin by considering the flat space Dirac
equation:}
\begin{equation}
{(i{\gamma}^{\mu}\,{\partial}_{\mu} - m)\,{\psi} = 0 .}
\end{equation}
{Everything we are about to say will go through for the curved space
version of eq. (3).
As is well-known, Dirac derived (3) by taking the square root of the standard
relativistic energy-momentum relation, and making the canonical substitutions of momenta
for differential operators: $p_{\mu} \,\rightarrow \, i\,{\partial}_{\mu}$. Dirac found
that the equation could only be satisfied if the ${\gamma}^{\mu}$s were actually $4
\times 4$ {\it matrices} satisfying precisely the Clifford algebra relation:}
\[
\{{\gamma}^{\mu}, {\gamma}^{\nu}\} = 2g^{\mu \nu} ,
\] 
{where $g^{\mu \nu}$ was (for Dirac) the flat Minkowski space metric. Thus, the actual
wavefunction ${\psi}$ representing the electron is a {\it four-component} object and
we are led naturally to the concept of antiparticles.

Once we form the set of solutions to equation (3) (and put an inner product structure
on that space so that it becomes a `Hilbert space', denoted ${\cal H}$), it is natural
to consider the representation of discrete geometrical transformations on ${\cal H}$.
Of paramount importance are the representations of $P$ (parity inversion) and 
$T$ (time reversal) which we must have if we are to construct a theory of 
elementary particles transforming under the action of the full 
inhomogeneous Lorentz group.

The best way to illustrate what we are talking about is with an explicit example. Let
us therefore recall how the operators $C$ (charge conjugation), $P$ (parity inversion)
and $T$ (time reversal) are represented in the standard particle physics literature
[5]: Let ${\cal H}$ be the set of solutions of the Dirac equation on four-dimensional
Minkowski space; then $C$, $P$, and $T$ are linear operators on ${\cal H}$ given by
the explicit formulae:}
\begin{eqnarray}
C: \,{\psi}(x, t) \longrightarrow i{\gamma}^{2}J{\psi}(x, t) , \nonumber \\
 \nonumber \\
P: \,{\psi}(x, t) \longrightarrow {\gamma}^{0}{\psi}(-x, t) , \\
 \nonumber \\
T: \,{\psi}(x, t) \longrightarrow {\gamma}^{1}{\gamma}^{3}J{\psi}(x, -t) , \nonumber
\end{eqnarray}
{where ${\psi}$ is any solution and J denotes the operation of 
complex conjugation. We remind the reader
that a host of physical considerations goes into the choices made in equations (4).
For example, the operator J is included in the construction of $T$ in order
to ensure that $T$ takes positive energy states to positive energy states.
A number of other choices are possible, the key point being that the other choices are
{\it mathematically inequivalent} (in a way to be made precise presently). 

Now, one of the first things we can notice about the operators $P$ and $T$ defined in
(4) is that they do not give a {\it Cliffordian} representation of the action of
space and time inversion. That is, 
$P$ and $T$ do not anti-commute, since in fact they commute:}
\[
PT \,\sim \, {\gamma}^{0}{\gamma}^{1}{\gamma}^{3} =
{\gamma}^{1}{\gamma}^{3}{\gamma}^{0} \,\sim \, TP .
\]
{Therefore, the operators $P$ and $T$ defined in (4) correspond to a {\it
non-Cliffordian} representation of $O(3, 1)$ with non-Cliffordian {\it action}.

This situation can be contrasted with the case where the representation is
Cliffordian. For example, a Cliffordian action can be recovered by the following
operator assignment:}
\begin{eqnarray}
P: \,{\psi}(x, t) \longrightarrow {\gamma}^{1}{\psi}(-x, t) , \nonumber \\
\\
T: \,{\psi}(x, t) \longrightarrow {\gamma}^{0}{\psi}(x, -t) . \nonumber
\end{eqnarray}
{Clearly, the choices in (5) anti-commute. 

Of course, in each of the above examples, the underlying group structure is identical.
More precisely, in the operator assignments made in (4), we used the group of elements
${\gamma}^{\mu}$ satisfying $\{{\gamma}^{\mu}, {\gamma}^{\nu}\} = 2g^{\mu \nu}$ to
construct operators $P$ and $T$ whose {\it action} on ${\cal H}$ is non-Cliffordian,
whereas in (5) we used the {\it same group} of Cliffordian elements to construct
operators $P$ and $T$ with {\it Cliffordian} action. It is absolutely essential that
we make this distinction between the different actions on a Hilbert space which can
be constructed from a given group, and genuinely {\it different groups}. This is
because we are sympathetic to the philosophy of Wigner [16] who put forward the idea
that the irreducible representations of whatever group of symmetries is 
present in nature
should form the basis for any theory of elementary particles. Indeed, Wigner
completely classified the set of irreducible representations of the inhomogeneous
Lorentz group, $O(3, 1)$, on the Hilbert space of solutions to the Dirac equation 
with $m \,\neq \, 0$. He showed that once one `fixes' the sign of the square of parity
inversion $P^{2}$ (fixing this sign corresponds to choosing a signature for spacetime,
basically) then there are four {\it inequivalent} (non-isomorphic) cases. The first
case is the standard particle physics choice made in (4) above. In the remaining three
cases, we encounter the phenomenon known as `parity doubling'.

Basically, then, there are eight different ways of representing the actions
of the operators $P$ and $T$ on the space of fermionic states.  Of course,
this should not surprise us too much since there are in fact eight distinct
non-isomorphic double covers of the inhomogeneous Lorentz group $O(p, q)$
when p and q are both non-zero.  Following Dabrowski,
we will write these covers as}
\[
{h^{a, b, c}:\, {\mbox{Pin}}^{a, b, c}(p, q) ~{\longrightarrow}~ O(p, q) , }
\]
{with $a, b, c ~{\in}~ {\{}+, -{\}}$ given as $a = P^{2}$, $b = T^{2}$, and 
$c = (PT)^{2}$.  Thus, a given double cover of $O(p, q)$ is completely
characterized by the signs of the squares of parity inversion, time reversal,
and the combination of the two.  These different double covers are called
the `pin' groups, and although our conventions for defining a, b, and c
differ from Dabrowski's (he takes $a = -(P)^{2}$), we feel that our 
notation (which is the notation used in [9]) makes the obstruction 
theory more transparent.

With this in mind, we can answer the obvious question: Which pin group corresponds to
the actions of $P$ and $T$ defined in equations (4) above? To see the answer, we
simply compute:}
\begin{eqnarray*}
P^{2} = ({\gamma}^{0})^{2} = - 1 , \\ 
 \\
T^{2} = ({\gamma}^{1}{\gamma}^{3})^{2} = - 1 , \\
 \\
(PT)^{2} = ({\gamma}^{0}{\gamma}^{1}{\gamma}^{3})^{2} = + 1 .
\end{eqnarray*}
{Thus, the pin group customarily used in particle physics is seen to be $\mbox{Pin}^{-,
-, +}(3, 1)$. This choice, we should point out, cannot be made flippantly since as was
pointed out in [18], a different choice for the representations of $P$ and $T$
corresponds to a different superselection sector of fermions, i.e., a completely
different species of particle. 

For example, let us consider a simple scattering experiment, where there is a bubble
of non-orientable foam.  Furthermore, let us assume that we obtain the bubble by 
some cut-and-paste
construction on Minkowski space, so that we can ignore any curvature effects and so
that the `$S$ matrix' describing scattering off the bubble is given simply as parity
inversion:}
\[
S = P
\]

{For example, we could simply decree that any causal path which intersects the 
spacelike surface ${\{}(x,t): |x| < 1, t = 0{\}}$ is parity reversing.}

{Thus, the operator representing $P$ now appears in the Hilbert space, since by
definition the final state ${\psi}_{f}$ is given in terms of the initial state 
${\psi}_{i}$ as}
\[
{\psi}_{f} = P\,{\psi}_{i}
\]
{($P$ is of course a unitary operator in the standard case, e.g. (4) above). It
follows that the solutions of the Dirac equation on this parity-reversing bubble are
actually sections of a {\it pin} bundle, i.e., a bundle whose fibres are isomorphic to
one of the above eight pin groups. Once we choose which representations of $P$ and $T$
we will employ, we determine the fibre group completely. Denote the chosen pin bundle
`${\cal B}$'. Then the only way in which we can sensibly operate on a section of 
${\cal B}$ is by using the fibres of ${\cal B}$, i.e., it is mathematically vacuous to
say that we want to consider the action on sections of ${\cal B}$ of a group which is
not isomorphic to the fibres of ${\cal B}$. Of course, there may exist other `types'
of fermions, given by different choices for $P$ and $T$, but they will not interact
with the fermions which lie in ${\cal B}$.

Let us now briefly return to our basic examples of non-orientable black holes, the
quotient manifolds ${\cal M}_{J}$ constructed above. Which of these manifolds admit 
$\mbox{Pin}^{-, -, +}(3, 1)$ structure? By the results of [9], we know that the
obstruction to $\mbox{Pin}^{-, -, +}(3, 1)$ structure is that the following
obstruction vanish on all two-cycles in ${\cal M}$:}
\[
w_{2} \,+\, w_{1}^{+} \,\smile\, w_{1}^{+} \,+\, w_{1}^{-} \,\smile\, w_{1}^{-}
\]
{where $w_{1}^{+}$ is the obstruction to space-orientability, $w_{1}^{-}$ is the
obstruction to time-orientability, $w_{2}$ is the second Stiefel-Whitney class, and
`$\smile$' denotes the cup product, as outlined in [9]. With these definitions in
mind, consider ${\cal M}_{P}$.

${\cal M}_{P}$ contains a single non-trivial two-cycle $c$, which is a spacelike
${\Bbb R}{\Bbb P}^{2}$. On this two-cycle, we therefore have}
\begin{eqnarray*}
w_{2}[c] = 1 \\ 
 \\
w_{1}^{+} \,\smile\, w_{1}^{+}[c] = 1 \\
 \\
w_{1}^{-} \,\smile\, w_{1}^{-}[c] = 0
\end{eqnarray*}
{where we are working in additive ${\Bbb Z}_{2}$. Thus, the above obstruction vanishes
$\bmod \,2$ and so ${\cal M}_{P}$ admits $\mbox{Pin}^{-, -, +}(3, 1)$ structure.
Proceeding in this vein for the other examples, we establish the following table:}
\begin{center}
\begin{tabular}{|l|c|}
\hline
 & \\
 &${{\cal M}_{J}}$ admits \\
\hspace*{0.9cm} $J$ &$\mbox{Pin}^{-, -, +}(3, 1)$ structure? \\
 & \\
\hline
 & \\
$J = R_{T}R_{Z}P$ &no  \\
 & \\
\hline
 & \\
$J = R_{T}P$ &yes \\
 & \\
\hline
 & \\
$J = R_{Z}P$ &yes  \\
 & \\
\hline
 & \\
$J = P$ &yes \\
 & \\
\hline
\end{tabular}
\end{center}
\begin{center}
{\large Table 2}
\end{center}
\vspace*{0.2cm}

{Of course, we are interested in matching these Lorentzian sections to their
Riemannian counterparts. We therefore want to know how to `Wick rotate' pinors. With
this in mind, let us first review the definition of pinors in Euclidean signature.

Really, the situation in Euclidean signature is much simpler: Given the orthogonal
group (inhomogeneous) in $n$-dimensions, $O(n)$, there are just {\it two}
double-covers of $O(n)$, usually denoted $\mbox{Pin}^{+}(n)$ and $\mbox{Pin}^{-}(n)$,
where the $\{\pm\}$ denotes the sign of the square of the element in
$\mbox{Pin}^{\pm}(n)$ which covers reflection in $O(n)$ [19]. The obstructions to
these structures are similar to the obstructions in Lorentzian signature, and can
be summarised as follows:}\\

{\noindent (1) There exists $\mbox{Pin}^{+}(n)$ structure iff $w_{2}(M) = 0$, i.e.,
iff the manifold is spin.}\\

{\noindent (2) There exists $\mbox{Pin}^{-}(n)$ structure iff $w_{2}(M) \,+\, w_{1}
\,\smile\, w_{1} = 0$, where $w_{1} \,\smile\, w_{1}$ is the cup product of the first
Stiefel-Whitney class of $M$ with itself.}\\

{Consider now the `identified cigar' instantons, the ${\cal M}_{J}^{R}$ constructed
above. As usual, let us start with ${\cal M}_{P}^{R}$. Originally, ${\cal M}^{R}$ is
topologically ${\Bbb R}^{2} \,\times\, S^{2}$. The identification under ${\bar P}$
corresponds to antipodal identification of the $S^{2}$ factor so that ${\cal
M}_{P}^{R}$ is topologically}
\[
{\cal M}_{P}^{R} \,\cong\, {\Bbb R}^{2} \,\times\, {\Bbb R}{\Bbb P}^{2}
\]
{${\cal M}_{P}^{R}$ is not spin, and so it does not admit $\mbox{Pin}^{+}(4)$
structure. On the other hand,}
\[
w_{1} \,\smile\, w_{1} [{\Bbb R}{\Bbb P}^{2}] = 1
\]
{and therefore ${\cal M}_{P}^{R}$ {\it does} admit $\mbox{Pin}^{-}(4)$ structure.
This is good, since if we want to `match' the $\mbox{Pin}^{-}(4)$ structure on ${\cal
M}_{P}^{R}$ to the $\mbox{Pin}^{-, -, +}(3, 1)$ structure on ${\cal M}_{P}$ (across
the Einstein Rosen bridge ${\Sigma}$, in a way to be made precise in a moment), then
we would want the signs of the squares of inversion to match as well. 

We then work out the obstructions to $\mbox{Pin}^{-}(4)$ on ${\cal M}_{J}$, for the
other values of ${\bar J}$, and obtain the table:}
\begin{center}
\begin{tabular}{|l|c|}
\hline
 & \\
 &${{\cal M}_{J}^{R}}$ admits \\
\hspace*{0.9cm} ${\bar J}$ &$\mbox{Pin}^{-}(4)$ structure? \\
 & \\
\hline
 & \\
${\bar J} = {\bar R_{T}}{\bar R_{Z}}{\bar P}$ &no  \\
 & \\
\hline
 & \\
${\bar J} = {\bar R_{T}}{\bar P}$ &yes \\
 & \\
\hline
 & \\
${\bar J} = {\bar R_{Z}}{\bar P}$ &yes  \\
 & \\
\hline
 & \\
${\bar J} = {\bar P}$ &yes \\
 & \\
\hline
\end{tabular}
\end{center}
\begin{center}
{\large Table 3}
\end{center}
\vspace*{0.2cm}

{Note how well Table 2 agrees with Table 3. Given this nice correspondence, we can now
describe how to `match' the pinors on ${\cal M}_{J}^{R}$ to the pinors on ${\cal
M}_{J}$ across ${\Sigma}$, so that the data induced on ${\Sigma}$ by the Euclidean
pinors agrees with the data induced on ${\Sigma}$ by the Lorentzian pinors.

To see how this works, first recall that the structure group of {\it complexified}
Schwarzschild ${\cal M}_{\Bbb C}$ is $SO(4, {\Bbb C})$ and that this group splits
naturally into two copies of $SL(s, {\Bbb C})$:}
\[
SO(4, {\Bbb C}) \,\simeq\, SL(2, {\Bbb C}) \,\times\, SL(2, {\Bbb C})
\]
{(Since the inclusion of inversions, i.e., passing from spin to pin, just involves
forming the semi-direct product of these groups with some finite discrete groups, we
really only need to show how to match the spinors on ordinary Schwarzschild ${\cal
M}^{L}$ across to the spinors on ${\cal M}^{R}$, and then our above discussion (Tables
2 and 3) will take care of the matching when one includes the involutions $J$).

Thus, we want to know how to match the $\mbox{Spin}(4)$ structure on ${\cal M}^{R}$ to
the $SL(2, {\Bbb C})$ structure on ${\cal M}^{L}$ across ${\Sigma}$. But that is easy.
After all,}
\[
\begin{array}{cccc}
\mbox{Spin}(4) ~~\,\simeq\, &SU(2) &\times &SU(2) \\
 &\cap & &\cap \\
 &SL(2, {\Bbb C}) &\times &SL(2, {\Bbb C}) 
\end{array}
\]
{and the $\mbox{Spin}(4)$ structure will induce $SU(2)$ spinors on the Einstein Rosen
bridge. We want to think of these spinors as `initial data' for the Lorentzian
section. But the spin structure on ${\cal M}^{L}$ is one of the $SL(2, {\Bbb C})$
factors in the above diagram. Therefore, in order to ensure that the two spin
structures match, we simply have to make sure that the structure group of 
${\cal M}^{L}$ is the $SL(2, {\Bbb C})$ factor which contains the $SU(2)$ factor
induced by the $\mbox{Spin}(4)$ structure on ${\cal M}^{R}$. 

{We can make these matching conditions more explicit by introducing
local coordinates for all of the spin structures involved. More precisely, at each point
$p \,\in\, {\cal M}_{\Bbb C}$, the tangent space is just}
\[
T_{p}({\cal M}_{\Bbb C}) \,\cong\, {\Bbb C}^{4}
\]
{As usual, we can do the `twistorial' thing [25] and rewrite ${\Bbb C}^{4}$ in terms of
$2 \,\times\, 2$ complex matrices, i.e., ${\Bbb C}^{4} \,\simeq\,  {\Bbb C}^{2 \,\times\,
2}$ and the isometry is just given by the map}
\[
{\Bbb C}^{4} \,\ni\, \left(z^{0}, z^{1}, z^{2}, z^{3}\right) \,\longrightarrow\, \left(
\begin{array}{cc}
z^{0} \,+\, z^{3} &z^{1} - iz^{2} \\
 & \\
z^{1} \,+\, iz^{2} &z^{0} \,+\, z^{3}
\end{array} \right)
= \left( z^{ij} \right) \,\in\, {\Bbb C}^{2 \,\times\, 2}
\]
{The fact that $SO(4, {\Bbb C})$ splits into a `left' and a `right' part (via 
$SO(4, {\Bbb C}) \,\simeq\, SL(2, {\Bbb C})_{L} \,\times\, SL(2, {\Bbb C})_{R}$) simply
means that the action of an element $(S_{L}, S_{R}) \,\in\, SO(4, {\Bbb C})$ on some
$\left( z^{ij} \right) \,\in\, {\Bbb C}^{2 \,\times\, 2}$ can be written}
\[
\left( z^{ij} \right) \,\longrightarrow\, S_{L}\left( z^{ij} \right) S_{R}^{-1}
\]
{Intuitively, what is going on is that at each point of ${\Bbb C}^{4}$ there are two
orthogonal two-dimensional complex planes, each of them acted on by an $SL(2, {\Bbb
C})$ (in a way reminiscent of the way ${\Bbb R}^{4}$ splits at each point into
orthogonal ${\Bbb R}^{2}$ factors, each associated with an $SU(2)$). ${\cal M}^{L}
\,\subset\, {\cal M}_{\Bbb C}$ has the property that at each $p$, $T_{p}({\cal M}^{L})
\,\simeq\, {\Bbb C}^{2}$ is acted upon by one of the $SL(2, {\Bbb C})$ factors which
we take to be $SL(2, {\Bbb C})_{L}$, without loss of generality. Thus, at each 
$T_{p}({\cal M}^{L})$ the above isometry becomes}
\[
{\Bbb R}^{3, 1} \,\ni\, (t, x, y, z) \,\longrightarrow\, \left( \begin{array}{cc}
t \,+\, z &x - iy \\
 & \\
x \,+\, iy &t - z
\end{array} \right)
\,\in\, H(2)
\]
{where $H(2) = \left\{ 2 \,\times\, 2 \;\mbox{Hermitian matrices} \right\}$. 
We can choose local Minkowskian coordinates about a point in the Einstein-Rosen bridge
so that  $t = 0$ corresponds to the intersection
${\cal M}^{L} \,\cap\, {\cal M}^{R} = \Sigma \,\not=\, \emptyset$, and so the above
Hermitian matrix reduces to}
\[
\left( \begin{array}{cc}
z &x - iy \\
 & \\
x \,+\, iy &- z
\end{array} \right)
\]
{which is an element of $SU(2)$, the spin group for $SO(3)$, i.e., the spatial
rotation group of $SL(2, {\Bbb C})_{L}$ is the $SU(2)$ induced on $\Sigma$ by the
Lorentzian section.}

We have now seen how to define fermions on the `classical' non-orientable black holes,
and we have touched on the thermodynamical properties of these objects. It is time we
turned to the problem of {\it creating} these sorts of objects using the instanton
approximation.}\\
\vspace*{0.6cm}

{\noindent \bf IV. Non-Orientable Instantons}\\

{We begin with the Schwarzschild-de Sitter solution, which may be
written in the following form:}
\begin{eqnarray}
ds^{2} = - \left(1 - \frac{2m}{r} - \frac{\Lambda r^{2}}{3}\right) dt^{2} \,&+&\, 
\left(1 - \frac{2m}{r} - \frac{\Lambda r^{2}}{3}\right)^{-1} dr^{2} \nonumber \\
 \nonumber \\
&+&\, r^{2}\left(d{\theta}^{2} \,+\, {\sin}^{2}{\theta}\,d{\phi}^{2}\right)
\end{eqnarray}
{As usual, we interpret this solution as  a black hole immersed in de Sitter
space. Upon Wick rotating this metric, one finds that there are apparent singularities
in the Riemannian section {\it unless} the cosmological constant ${\Lambda}$ and the
mass $m$ of the hole are related by}
\[
\sqrt{\Lambda} = \frac{1}{3m}
\]
{This equality corresponds to the limit in which the black hole and cosmological event
horizons merge. Using this equality, the metric on the Riemannian section becomes}
\[
ds^{2} = \left(1 - \Lambda {\rho}^{2}\right) d{\tau}^{2} \,+\, \frac{d {\rho}^{2}}{1 -
\Lambda {\rho}^{2}} \,+\, \frac{1}{\Lambda}\left(d{\theta}^{2} \,+\, 
{\sin}^{2}{\theta}\,d{\phi}^{2}\right)
\]
{To see that this metric lies on $S^{2} \,\times\, S^{2}$, one introduces the
coordinate ${\eta}$ via}
\[
\rho \sqrt{\Lambda} = \cos \eta
\]
{whence the metric becomes}
\begin{equation}
ds^{2} = \frac{1}{\Lambda}\left(d{\eta}^{2} \,+\, {\sin}^{2}{\eta}\,
d{\tau}^{2} \,+\, d{\theta}^{2} \,+\, {\sin}^{2}{\theta}\,d{\phi}^{2}\right)
\end{equation}

{The Euclidean action is calculated to be}
\[
{\cal S} = \frac{-2\pi}{\Lambda}
\]
{The metric on (7) is manifestly the standard product metric on $S^{2} \,\times\,
S^{2}$, where each $S$ has radius $\frac{1}{\sqrt{\Lambda}}$.

We now wish to form the quotient of this instanton by some discrete involutive freely
acting isometries. There will be a number of different possibilities for the set of
maps under which we can identify. Let us begin with the Lorentzian section (6). 

First, we transform to coordinates $(t, \chi, \theta, \phi)$, where the metric takes
the form}
\begin{equation}
ds^{2} = \frac{1}{\Lambda}\left(- dt^{2} \,+\, \left({\cosh}\left(\sqrt{\Lambda}
\,t\right)\right)^{2}\,
d{\chi}^{2} \,+\, d{\theta}^{2} \,+\, {\sin}^{2}{\theta}\,d{\phi}^{2}\right)
\end{equation}
{i.e., the coordinates $t$ and $\chi$ are (respectively) timelike and spacelike
coordinates on two-dimensional de Sitter $(- \infty < t < \infty , \, 0 \leq \chi \leq
2\pi)$ and $(\theta , \phi)$ are the usual coordinates on $S^{2}$ $(0 \leq \theta \leq
\pi, \, 0 \leq \phi \leq 2\pi)$. In terms of these coordinates, there are then several
involutions which we will make use of.

First of all, there is time reversal:}
\[
T: \, (t, \chi , \theta , \phi) \,\longrightarrow\, (- t, \chi , \theta , \phi)
\]
{This obviously has fixed points $(t = 0)$.

Next, there is inversion in the spacelike coordinate:}
\[
I: \, (t, \chi , \theta , \phi) \,\longrightarrow\, (t, \chi + \pi , \theta , \phi)
\]
{Intriguingly, this involution has no fixed points!

Finally, there is the usual freely acting involution of parity:}
\[
P: \, (t, \chi , \theta , \phi) \,\longrightarrow\, (t, \chi , \pi - \theta , \phi 
+ \pi )
\]
{Therefore, on the Lorentzian section at least, we
now have {\it two} freely acting isometries, and thus a much richer
range of possibilities. Let $M$ denote the Schwarzschild-de Sitter solution (without
identifications). Then we shall follow the notation of Section 2 above when we form
quotient spaces, i.e.,}
\[
M_{P} = {M}/{P} , M_{I} = {M}/{I} , ~~~\mbox{etc.}
\]

{As an amusing aside, consider the spacelike $M_{TI}$. This is a non-time orientable
spacetime with a single boundary component 
homeomorphic to $S^{1} \,\times\, S^{2}$, and in fact the manifold is
basically two-dimensional antipodally identified de Sitter crossed with a two-sphere
(see [13], [14] for more on antipodally identified de Sitter). Thus, this spacetime
can be thought of as a {\it Lorentzian path} corresponding to the birth of a universe
(with a black hole in it) from nothing. We have more about the interplay between
Lorentzian path integrals and Euclidean instantons in another recent paper [21].

Now, in analogy with what we did above in Sec. I, let us consider the
Riemannian section $M^{R} \,\cong\, S^{2} \,\times\, S^{2}$, and let's construct
the maps ${\bar T}$, ${\bar I}$ and ${\bar P}$ which correspond to the maps
$T$, $I$ and $P$ on the Lorentzian section.  As was done above, we could
embed both the Lorentzian and Riemannian sections in some higher 
dimensional complex space (the Riemannian section would then just be
a product of complex projective lines), and we could find `big maps' on
the higher dimensional complex manifold which yielded the desired
involutions when restricted to the real or imaginary time sections.
However, the geometry in this situation is so simple that we can just 
write down the  involutions on the Riemannian instanton by
inspection.  In terms of the coordinates used in equation (8) above,
and remembering to change to imaginary time ${\tau} = it$, 
these involutions are given as follows:}
\[
{\bar T}: \, ({\tau}, \chi , \theta , \phi) \,\longrightarrow\, (-{\tau}, \chi , \theta , \phi)
\]
{(${\tau}$ now has range $- {\pi}/2 < {\tau} < {\pi}/2$).}
\[
{\bar I}: \, ({\tau}, \chi , \theta , \phi) \,\longrightarrow\, ({\tau}, \chi + \pi , \theta , \phi)
\]
{And as usual, parity:}
\[
{\bar P}: \, (t, \chi , \theta , \phi) \,\longrightarrow\, (t, \chi , \pi - \theta , \phi 
+ \pi )
\]
{Thus, we see that the transition from the Lorentzian solution to the 
Riemannian instanton is rather simple in this example.  Now notice, however,
that ${\bar I}$ is ${\it not}$ freely acting on the $({\tau}, {\chi})$ sphere
(it has fixed points at the north and south poles) and so we cannot use
${\bar I}$ to construct freely acting involutions on the instanton.  There are
thus four instantons which describe the birth from nothing of a pair of
non-orientable black holes in a de Sitter background (or, equivalently,
one could use these instantons to calculate the rate of decay of de Sitter
space into such a black hole pair).  These instantons are (using the usual
notation): $M^{R}_{P}$, $M^{R}_{IP}$, $M^{R}_{TP}$, and $M^{R}_{TIP}$.
As usual, we summarize the properties of the Lorentzian solutions 
which these instantons correspond to in a table:
\begin{center}
\begin{tabular}{|l|c|c|c|}
\hline
 & & & \\
 &${{M}_{J}}$ &${{M}_{J}}$ 
&${{M}_{J}}$ \\
\hspace*{1.5cm} $J$ &admits $Pin^{-,-,+}(3,1)$ &time &space \\
 &structure? &orientable? &orientable? \\
 & & & \\
\hline
 & & & \\
Case I: $J = TIP$ &no &no &yes \\
 & & & \\
\hline
 & & & \\
Case II: $J = TP$ &yes &no &no \\
 & & & \\
\hline
 & & & \\
Case III: $J = IP$ &yes &yes &yes \\
 & & & \\
\hline
 & & & \\
Case IV: $J = P$ &yes &yes &no \\
 & & & \\
\hline
\end{tabular}
\end{center}
\begin{center}
{\large Table 4}
\end{center}
\vspace*{0.2cm}
 
{\noindent As discussed above, the probability for de Sitter space
to decay into one of these non-orientable Lorentzian solutions is given 
by the square of the amplitude, where the amplitude is given in the semiclassical 
approximation by $e^{-S}$ with $S$ the Euclidean action.  (Actually, we have
to first divide the amplitude to create a universe with black holes by the 
amplitude to create ordinary de Sitter in order to obtain the rate of decay of
de Sitter, but we will overlook that subtlety here).  Consider the instanton}
\[
{M}^{R}_{P} \,\cong\, {S}^{2} \,\times\, {\Bbb R}{\Bbb P}^{2}
\]
{We would like to know whether de Sitter is more likely to decay into
$M$ or $M_P$.  As we noted above, the action for $M^R$ is given as
${\cal S} = \frac{-2\pi}{\Lambda}$.  On the other hand, the Euler number of
$M^{R}_{P}$ is 2 (half that of $M^R$) and so the action ${\cal S}_P$ must be
${\cal S}_{P} = \frac{-\pi}{\Lambda}$.  Thus, recalling that we actually only
need the action for `half' the instanton and that the probability is the square of
the resulting amplitude, it would seem that the 
probabilities for these two decay processes differ by a factor of $e^{{\pi}/{\Lambda}}$,
which measures the suppression of the rate of non-orientable hole production
relative to the rate of orientable hole production.}

{Of course, the specific results which we have outlined here will go through
in general for any instanton and corresponding classical solution which
admit at least one discrete, freely acting isometry.  For example, the 
Mellor-Moss instanton [22] (describing the nucleation of charged black holes
in a de Sitter background) admits a freely acting isometry as do all of the 
instantons obtained from the many solutions, derived from the C-metric,
which describe the production of charged black holes in background 
fields [23].  And we have also recently pointed out [21] that non-orientable
black holes will be created in the presence of vacuum domain walls.
It therefore seems that whenever one has an energy source which can
contribute to tunneling phenomena corresponding to the birth of ordinary
black holes, that same energy source will also contribute to non-orientable
black hole pair production.  The only caveat is that the rate of non-orientable
black hole production will be supressed relative to the rate of 
production of ordinary holes, since the identified instantons will generically
have less volume (and hence less action)  than the original orientable 
instantons.  }
\vspace*{0.6cm}

{\noindent \bf Acknowledgements}\\

{The authors  would like to thank J. Louko for a useful discussion.
Thanks also go to Jo Chamblin (Piglit) for
help with the preparation of this paper, and to R. Bousso for pointing out typos in
an earlier draft.  A.C.
was supported by NSF Graduate Fellowship No. RCD-9255644 (in Cambridge) and 
NSF PHY94-07194 (in Santa Barbara). }\\
\vspace*{0.6cm}

{\noindent \bf References}\\

{\noindent [1] S.W. Hawking, {\it Virtual Black Holes}, DAMTP Preprint \#R95/
(1995).}\\

{\noindent [2] G.W. Gibbons, in {\it Fields and Geometry 1986}, Proceedings of the
22nd Karpacz Winter School of Theoretical Physics, Karpacz, Poland, ed. A. Jadczyk,
World Scientific, Singapore (1986).}\\

{\noindent [3] M.J. Perry, {\it An Instability of De Sitter Space}, in {\it The Very
Early Universe}, ed. G.W. Gibbons, S.W. Hawking and S.T.C. Siklos, CUP (1983).}\\

{\noindent [4] P. Ginsbarg and M.J. Perry, Nucl. Phys. {\bf B 222}, 245 (1983).}\\

{\noindent [5] J.D. Bjorken and S.D. Drell, {\it Relativistic Quantum Mechanics and
Relativistic Quantum Fields}, McGraw-Hill, New York (1965).}\\

{\noindent [6] G. 't Hooft, J. Goem. Phys. {\bf 1}, 45 (1984).}\\

{\noindent [7] G.W. Gibbons, {\it The Elliptic Interpretation of Black Holes and
Quantum Mechanics}, Nucl. Phys. {\bf B 271}, 497--508 (1986).}\\

{\noindent [8] G.W. Gibbons, {\it Topology Change in Classical and Quantum Gravity},
Proc. of 10th Sorak School of Theor. Phys., ed. J.E. Kim, World Scientific (1992).}\\

{\noindent [9] A. Chamblin, {\it On the Obstructions to non-Cliffordian Pin
Structures}, Comm. Math. Phys. {\bf 164}, No. 1, 65--87 (1994).}\\

{\noindent [10] W. Israel, Phys. Lett. {\bf 57 A}, 107 (1976).}\\

{\noindent [11] W. Israel, Phys. Rev. Lett. {\bf 155}, 1001 (1965).}\\

{\noindent [12] M.D. Kruskal, {\it Maximal Extension of Schwarzschild Metric}, Phys.
Rev. {\bf 119}, 1743--1745 (1960).}\\

{\noindent [13] J.L. Friedman, {\it Two-component spinor fields in a class of time-
nonorientable spacetimes}, Class. Quant. Grav., Vol. 12, No. 9, 2231-2242 (1995).}\\

{\noindent [14] A. Chamblin and G.W. Gibbons, {\it A judgement on sinors}, Class. 
Quant. Grav., Vol. 12, No. 9, 2243-2248 (1995).}\\

{\noindent [15] G.W. Gibbons and S.W. Hawking, {\it Selection Rules for Topology
Change}, Comm. Math. Phys. {\bf 148}, page 345 (1992).}\\

{\noindent [16] E.P. Wigner, {\it Group Theoretical Concepts and Methods in
Elementary Particle Physics}, (ed. F. Gursey), Gordon and Breach (1962).}\\

{\noindent [17] L. Dabrowski, {\it Group Actions on Spinors}, Monographs and
Textbooks in Physical Science, Bibliopolis (1988).}\\

{\noindent [18] D. Ebner, {\it Classification of Relativistic Particles According to
the Representation Theory of the Eight Non-isomorphic Simply Connected Covering
Groups of the Full Lorentz Group}, Gen. Rel. Grav. {\bf 8}, No. 1, 15--28 (1977).}\\

{\noindent [19] R.C. Kirby and L.R. Taylor, {\it Pin Structures on Low-Dimensional
Manifolds}, London Math. Soc. Lecture Notes, No. 151, CUP (1989).}\\

{\noindent [20] G.W. Gibbons and M.J. Perry, {\it Black Holes and Thermal Green
Functions}, Proc. Roy. Soc. Lond. A {\bf 358}, 467--494 (1978).}\\

{\noindent [21] R. Caldwell, A. Chamblin and G.W. Gibbons, {\it Pair Creation of
Black Holes in the Presence of Domain Walls}, Phys. Rev. {\bf D 53}, 7103 (1996).}\\

{\noindent [22] F. Mellor and I. Moss, {\it Black holes and quantum wormholes},
Phys. Lett. {\bf 222B}, 361-363 (1989).}\\

{\noindent [23] W. Kinnersley and M. Walker, Phys. Rev. {\bf D 2}, 1359 (1970); 
F.J. Ernst, J. Math. Phys. {\bf 17}, 515 (1976)}\\

{\noindent [24] H. Seifert and W. Threlfall, {\it A Textbook of Topology}, Chapter IX,
Theorem IV, Academic Press (1980).}\\

{\noindent [25] R.S. Ward and R.O. Wells, {\it Twistor Geometry and Field Theory},
Cambridge University Press, Cambridge (1990).

\end{document}